\documentclass[aps,prl,twocolumn,superscriptaddress,bibnotes]{revtex4-1}
\pdfoutput=1

\usepackage{graphicx}
\usepackage{bm}
\usepackage{amssymb,amsmath,latexsym}
\usepackage{color}
\definecolor{darkblue}{RGB}{0,0,196}
\definecolor{darkred}{RGB}{196,0,0}
\usepackage[colorlinks=true,linktocpage=true,linkcolor=darkblue,citecolor=darkblue,urlcolor=darkblue]{hyperref}
\usepackage{array}
\usepackage{soul}

\begin{document}

\title{The impact of fluctuating initial conditions on bottomonium suppression in 5.02 TeV heavy-ion collisions}

\author{Huda Alalawi} 
%\email{halalawi@kent.edu}
\affiliation{Department of Physics, Kent State University, Kent, OH 44242, USA}

\author{Jacob Boyd}
%\email{jboyd29@kent.edu}
\affiliation{Department of Physics, Kent State University, Kent, OH 44242, USA}

\author{Chun Shen}
%\email{chunshen@wayne.edu}
\affiliation{Department of Physics and Astronomy, Wayne State University, Detroit, Michigan, 48201, USA}
\affiliation{RIKEN BNL Research Center, Brookhaven National Laboratory, Upton, NY 11973, USA}

\author{Michael Strickland} 
%\email{mstrick6@kent.edu}
\affiliation{Department of Physics, Kent State University, Kent, OH 44242, USA}

\begin{abstract}
We compute bottomonium suppression and elliptic flow within the pNRQCD effective field theory using an open quantum systems approach.
For the hydrodynamical background, we use 2+1D MUSIC second-order viscous hydrodynamics with IP-Glasma initial conditions and evolve bottom/antibottom quantum wave packets in real time in these backgrounds.
We find that the impact of fluctuating initial conditions is small when compared to results obtained using smooth initial conditions.  Including the effect of fluctuating initial conditions, we find that the $\Upsilon(1S)$ integrated elliptic flow is \mbox{$v_2[1S] = 0.005 \pm 0.002 \pm 0.001$}, with the first and second variations corresponding to statistical and systematic theoretical uncertainties, respectively.  
\end{abstract}

\date{\today}


\maketitle

The strong suppression of bottomonium production in heavy-ion collisions relative to their production in proton-proton collisions is a smoking gun for the creation of a hot quark-gluon plasma (QGP) in relativistic heavy-ion collisions~\cite{Acharya:2020kls,ATLAS5TeV,Sirunyan:2018nsz,CMS:2017ycw,CMS:2020efs,ALICE:2019pox,STAR:2013kwk,PHENIX:2014tbe,STAR:2016pof,CMSupsilonQM2022}.
In the seminal work of Matsui and Satz \cite{Matsui:1986dk}, suppression of heavy quarkonium was proposed as a signal of the formation of a color-ionized QGP.  It was conjectured to result from Debye screening of chromoelectric fields in a QGP, which causes a modification of the heavy-quark heavy-anti-quark potential.  

This turned out to be only partly true.  In recent years it was shown that, in addition to Debye screening of the real part of the potential, there also exists an imaginary contribution to the  potential, which results in large in-medium widths for heavy quarkonium bound states~\cite{Laine:2006ns,Brambilla:2008cx,Beraudo:2007ky,Escobedo:2008sy,Dumitru:2009fy,Brambilla:2010vq,Brambilla:2011sg,Brambilla:2013dpa}.  
The existence of an imaginary part of the potential results from two main effects: Landau damping, which is related to parton dissociation \cite{Brambilla:2013dpa}, and singlet to octet transitions, which is an effect particular to QCD \cite{Brambilla:2008cx} related to the gluo-dissociation process \cite{Brambilla:2011sg}.  Including the imaginary part of the potential in calculations results in bottomonium states having widths on the order of 10-100 MeV in the QGP.
At temperatures relevant in current heavy-ion collision experiments, the presence of a large imaginary part of the potential is the most important cause of in-medium quarkonium suppression.
 
Resummed perturbative and effective field theory calculations of the imaginary part of the heavy quark potential have recently been confirmed by non-perturbative lattice QCD and classical real-time lattice measurements of the imaginary part of the potential~\cite{Rothkopf:2011db,Rothkopf:2019ipj,Petreczky:2018xuh,Bala:2019cqu,Laine:2007qy,Lehmann:2020fjt,Boguslavski:2020bxt} and complex potential models have been quite successful phenomenologically~\cite{Strickland:2011mw,Strickland:2011aa,Krouppa:2015yoa,Islam:2020bnp,Islam:2020gdv,Brambilla:2020qwo,Brambilla:2021wkt,Brambilla:2022ynh,Wen:2022yjx}.  These studies have provided strong evidence that a self-consistent quantum mechanical description of heavy quarkonium propagation in the QGP is possible.

The formalism used in this work is based on recent advances in our understanding of non-relativistic effective field theory (EFT) and real-time evolution in open quantum systems (OQS)~\cite{Breuer:2002pc}.
Such descriptions can model both screening effects and in-medium dynamical transitions between different color and angular momentum states.  
Recently, there has been a great deal of work on the application of OQS methods to heavy-quarkonium suppression \cite{Akamatsu:2011se,Akamatsu:2020ypb,Akamatsu:2014qsa,Miura:2019ssi,Brambilla:2016wgg,Brambilla:2017zei,Brambilla:2019tpt,Sharma:2019xum,Blaizot:2015hya,Blaizot:2017ypk,Blaizot:2018oev,Blaizot:2021xqa,Yao:2018nmy,Yao:2020xzw,Yao:2020eqy,Yao:2021lus,Katz:2015qja}.

Herein, we will apply OQS methods within the framework of the potential non-relativistic QCD (pNRQCD) EFT \cite{Pineda:1997bj,Brambilla:1999xf,Brambilla:2004jw}, which can be obtained systematically from non-relativistic QCD~\cite{Caswell:1985ui,Bodwin:1994jh}, and ultimately QCD itself.
The pNRQCD EFT relies on there being a large separation between the energy scales in the problem, which is guaranteed for systems where the velocity of the heavy quark relative to the center of mass is small, i.e., $v \ll 1$.
Such EFTs can also be used to study quarkonium at finite temperature \cite{Brambilla:2008cx,Escobedo:2008sy,Brambilla:2010vq,Brambilla:2011sg,Brambilla:2013dpa}.
Here we will assume the following scale hierarchy, which is appropriate at high temperatures: $1/r \gg m_D \sim \pi T \gg E$, where $r$ is the typical size of the state, $m_D$ is the Debye screening mass, $T$ is the local QGP temperature, and $E$ is the binding energy of the state.

In Refs.~\cite{Brambilla:2016wgg,Brambilla:2017zei,Brambilla:2019tpt} the authors derived a Lindblad equation \cite{Gorini:1975nb,Lindblad:1975ef} for the heavy quarkonium reduced density matrix using the scale hierarchy above. In the last two years, it has become possible to solve the Lindblad equation obtained numerically~\cite{Omar:2021kra}.  
The resulting code, called {\tt QTraj}, relies on a Monte Carlo quantum trajectories algorithm \cite{Dalibard:1992zz} to make phenomenological predictions~\cite{Brambilla:2020qwo,Brambilla:2021wkt,Brambilla:2022ynh}.
For this purpose, the Lindblad solver was coupled to a 3+1D viscous hydrodynamics code that used smooth (optical) Glauber initial conditions \cite{Alqahtani:2020paa,Alalawi:2021jwn}.  
The authors found that this provided a quite reasonable description of existing experimental data for both the nuclear modification factor, $R_{AA}$, and elliptic flow, $v_2$.  
It was also found in Refs.~\cite{Brambilla:2020qwo,Brambilla:2021wkt,Brambilla:2022ynh} that, to a very good approximation, one can compute the survival probability of quarkonium states by ignoring the off-diagonal jump terms that change the quantum numbers of the state, evolving instead with a self-consistently determined complex Hamiltonian for singlet states.

Herein, we make the first study of the effect fluctuations in the initial geometry have on bottomonium suppression and flow using an OQS framework that includes the real-time evolution of the bottomonium wave function in a complex potential.  Also for the first time, for the hydrodynamic background of the quantum evolution, we make use of the MUSIC hydrodynamics package with fluctuating IP-Glasma initial conditions.  This framework successfully describes a wide range of soft-hadron observables in heavy-ion collisions, e.g., identified particle spectra and anisotropic flow coefficients $v_n$~\cite{McDonald:2016vlt, Schenke:2020mbo}.
The IP-Glasma initial conditions incorporate gluon saturation effects in the initial state \cite{Bartels:2002cj,Kowalski:2003hm,Schenke:2012wb} and allow one to faithfully describe the early-time dynamics of the QGP.

Previous studies of the impact of fluctuations on bottomonium suppression were reported in Refs.~\cite{Song:2011ev,Song:iopproc}, where comparisons between smooth Glauber and Monte-Carlo Glauber initial conditions were made with the underlying model for quarkonium dynamics being eikonal transport with a space-time dependent disassociation rate.  Recently, in Ref.~\cite{Kim:2022lgu} the authors made use of Monte-Carlo Glauber initial conditions and the SONIC hydrodynamics code~\cite{SONIC} to make predictions for bottomonium production in pp, pA, and AA collisions.  In Ref.~\cite{Kim:2022lgu}, $R_{AA}$ was computed in the adiabatic approximation, by incorporating a temperature- and $p_T$-dependent thermal width, which included both gluo-dissocation and inelastic parton scattering~\cite{Hong:2019ade}. Finally, in Ref.~\cite{Yao:2020xzw} the authors considered fluctuating initial conditions in the context of a OQS-derived transport model. 

Herein, we will focus on AA collisions, but go beyond the adiabatic approximation and transport models by solving for the real-time quantum mechanical evolution along each sampled bottomonium trajectory.  This is particularly important at early times when the temperature depends strongly on proper time and, due to the inclusion of fluctuations, the position of the bottomonium state in the plasma.  The use of real-time quantum evolution also allows us to take into quantum state mixing due to the time-dependent potential energy.

\vspace{1mm}

{\em Methodology} --- To compute the survival probabilities of various bottomonium states, we evolve the bottom-antibottom wave function forward in time using a time-dependent complex effective Hamiltonian that is accurate to next-to-leading order (NLO) in the binding energy over temperature \cite{Akamatsu:2020ypb,Brambilla:2022ynh}. It can be expressed in terms of two parameters, $\kappa$ and $\gamma$, that can be obtained from the imaginary and real parts of a time-ordered correlator of chromo-electric fields, respectively.  The parameters $\kappa$ and $\gamma$ set the magnitude of the decay widths of the states and their mass shifts, respectively. 

When expressed as operators acting on the reduced wave function, $u = r R(r)$, the NLO singlet effective Hamiltonian $H^{\rm eff}_s$ is given by~\cite{Brambilla:2022ynh}
\begin{align}
    \text{Re}[H^{\rm eff}_s] &=  
    \frac{\nabla^2}{M}
    - \frac{C_F\, \alpha_{\rm s}}{r} + \frac{\hat\gamma T^3}{2} r^2 +  \frac{\hat\kappa T^2}{4 M} \{r,p_r\} \, , \label{eq:heff1} \\
    \text{Im}[H^{\rm eff}_s] &= - \frac{\hat\kappa T^3}{2} \Bigg[ \left( r - \frac{N_c \alpha_{\rm s}}{8T} \right)^2 - \frac{3}{2MT} \nonumber \\
    & \hspace{1.2cm} + \frac{\nabla^2}{(2MT)^2} + \frac{1}{2MT} \left( \frac{N_c \alpha_{\rm s}}{4 T} \right)  \frac{1}{r}\Bigg] ,
\end{align}
where $M$ is the heavy quark mass, $T$ is the temperature of the QGP, $N_c$ is the number of colors, \mbox{$C_F = (N_c^2-1)/2N_c$}, $\alpha_s$ is the strong coupling, \mbox{$p_r = - i \partial_r$}, and $\nabla^2 = -\partial_r^2 + l(l+1)/r^2$.  The rescaled transport coefficients appearing above are $\hat{\kappa} = \kappa / T^3$ and  $\hat{\gamma} = \gamma / T^3$.  The values of these coefficients were taken from direct and indirect lattice measurements and the uncertainty bands and central values used herein are the same as those used in Ref.~\cite{Brambilla:2022ynh}.  We take the heavy quark mass to be given by $M = M_{1S} = 4.73$ GeV and the strong coupling is set at the scale of the inverse Bohr radius to be \mbox{$\alpha_s(1/a_0) = 0.468$}.

We evolved ensembles of bottom/anti-bottom quantum wave packets with this effective Hamiltonian and ignore the effects of off-diagonal quantum jumps.  This has been shown to be a very good approximation in QCD by prior works \cite{Brambilla:2020qwo,Brambilla:2021wkt,Brambilla:2022ynh}.  To solve for the real-time evolution of each quantum wave packet, we used the Crank--Nicolson method.  We employed a one-dimensional lattice with $\texttt{NUM}=2048$ points and $\texttt{L}=40\,\mathrm{GeV}^{-1}$. The temporal step size was taken to be $\texttt{dt}=0.001\,\mathrm{GeV}^{-1}$~\footnote{For a systematic study of the lattice spacing and step size dependence, we refer the reader to Ref.~\cite{Omar:2021kra}.}.
The input for the quantum evolution necessary was the temperature experienced by the bottomonium wave packet along its trajectory through the QGP.

For the background QGP evolution, we considered \mbox{5.02 TeV} Pb-Pb collisions modeled by the MUSIC viscous hydrodynamics package \cite{Schenke:2010nt,Schenke:2011bn,MUSIC}, which includes the effects of both shear and bulk viscosity \cite{Ryu:2015vwa, Paquet:2015lta}.  For the hydrodynamic initial conditions, we use IP-Glasma fluctuating initial conditions which incorporate the effects of the dense gluonic environment generated at early-times in a nucleus-nucleus collision \cite{Bartels:2002cj,Kowalski:2003hm,Schenke:2012wb}.  For this work, we considered 2+1D boost-invariant evolution using MUSIC with a box size of $L=30$~fm and 512 grid points in both the $x$ and $y$ directions.  The equation of state used was based on the HotQCD lattice result \cite{Bazavov:2009zn, HotQCD:2014kol}. The shear viscosity to entropy density ratio used was $\eta/s=0.12$ and we made use of a temperature-dependent bulk viscosity~\cite{Schenke:2020mbo}.  With these parameters, the MUSIC hydrodynamics code is able to well describe, e.g., charged particle multiplicities, identified hadron spectra, and identified hadron anisotropic flow coefficients~\cite{Schenke:2020mbo}.

The ensembles of hydrodynamic events with IP-Glasma fluctuating initial conditions are sorted by the final charged hadron multiplicity into each of the following centrality bins 0-5\%, 5-10\%, 10-20\%, 20-30\%, 30-40\%, 40-50\%, 50-60\%, 60-70\%, 70-80\%, 80-90\%, and 90-100\%, with approximately 100 sampled IP-Glasma events in the 0-5\% and 5-10\% bins and 200 sampled events in the other centrality bins~\footnote{The evolution files used are publicly available on Google Drive, \url{https://drive.google.com/drive/folders/1rEF7Cfe2DMHmZmlUyGvMW4RjM5BijqMt?usp=share_link}}. In each centrality bin, we initialized the bottomonium evolution by sampling the initial production points from the fluctuating initial binary collision profile, which automatically includes correlations with the local hot spots generated in each event.  The initial transverse momentum was sampled from a $1/E_T^4$ spectrum and the azimuthal angle of the momentum direction was sampled uniformly in $[0,2\pi)$.  We assumed that the created bottom-antibottom wave packets traveled along eikonal trajectories with fixed transverse momentum and azimuthal angle and sampled the temperature along each trajectory generated.  Due to the fluctuating nature of initial conditions used in the hydrodynamical simulation, the temperature along each trajectory can be a strongly varying function of time.  For all results presented herein, we sampled 200,000 wave packet trajectories.

To initialize the real-time quantum evolution along each sampled trajectory, we assumed that at $\tau = 0$ fm the wave function was a localized delta function centered at the sampled production point and that the system was in the singlet state, with either $l=0$ or $l=1$ as the angular momentum quantum number.  We took the initial reduced wave function $u$ to be given by a Gaussian multiplied by a power of $r$ appropriate for the angular momentum of the state $l$,
$u_{\ell}(t_0) \propto r^{l+1} e^{-r^{2}/(ca_{0})^2}$,
with $u$ normalized to one and $c=0.2$ following earlier works~\cite{Omar:2021kra} \footnote{Observables do not depend significantly on $c$ below this value~\cite{Brambilla:2020qwo,Brambilla:2021wkt,Omar:2021kra,Brambilla:2022ynh}.}.
We evolved the initial wave function using the vacuum potential from $\tau = 0$~fm to $\tau_{\rm med}$ = 0.6~fm and we evolved the wave function with the vacuum potential whenever the temperature along the trajectory dropped below $T_F = 190$ MeV. This lower temperature cutoff was fixed by analyzing the convergence of the singlet width when going from LO to NLO in $E/T$ in Ref.~\cite{Brambilla:2022ynh}.  

At the end of the evolution along each physical trajectory we computed the survival probability of each of the vacuum eigenstates by projecting the final time-evolved wave function with vacuum bottomonium eigenstates, corresponding to the 1S, 2S, 3S, 1P, 2P, and 1D states.  In order to compare to data it is necessary to take into account late time feed down of the excited states.  Following Ref.~\cite{Brambilla:2020qwo}, this was accomplished using a feed down matrix $F$ that relates the experimentally observed (post feed down) and direct production cross sections (pre feed down) cross sections, $\vec{\sigma}_{\text{exp}} = F \vec{\sigma}_{\text{direct}}$.

The cross section vectors correspond to the states considered, while $F$ is a matrix, the values of which were fixed by the branching fractions of the excited states.
In our analysis, the states considered were $\vec{\sigma} = \{ \Upsilon(1S),\,$ $\Upsilon(2S),\,$ $\chi_{b0}(1P),\,$ $\chi_{b1}(1P),\,$ $\chi_{b2}(1P),\,$ $\Upsilon(3S),\,$ $\chi_{b0}(2P),\,$ $\chi_{b1}(2P),\,$ $\chi_{b2}(2P)\}$.
The entries of $F$ are $F_{ij} = B_{j \rightarrow i}$ for $i<j$, $F_{ij}$ = 1 for $i=j$, and $F_{ij}$ = 0 for $i>j$, where $B_{j \rightarrow i}$ is the branching fraction of state $j$ into state $i$.
The branching fractions were taken from the Particle Data Group listings~\cite{Zyla:2020zbs}.  We used the same branching fractions as prior OQS+pNRQCD papers, which used smooth hydrodynamic backgrounds~\cite{Brambilla:2020qwo,Brambilla:2021wkt,Brambilla:2022ynh}.  The entries of $F$ can be found in Eq.~(6.4) of Ref.~\cite{Brambilla:2020qwo}.

Finally, the nuclear modification factor $R_{AA}^i$ for bottomonium state $i$ is given by
\begin{equation}
R^{i}_{AA}(c,p_T,\phi) = \Bigg\langle\!\!\!\!\Bigg\langle \frac{\left(F \cdot S(c,p_T,\phi) \cdot \vec{\sigma}_{\text{direct}}\right)^{i}}{\vec{\sigma}_{\text{exp}}^{i}} \Bigg\rangle\!\!\!\!\Bigg\rangle \, ,
\label{eq:feeddown}
\end{equation}
where $i$ labels the bottomonium state being considered, $S(c,p_T,\phi)$ is the survival probability computed from the real-time quantum mechanical evolution, $c$ labels the event centrality class, $p_T$ is the transverse momentum of the bottomonium state, and $\phi$ its azimuthal angle. The angle brackets indicate a double average over (1) all physical trajectories of bottomonium states in the centrality and $p_T$ bin considered and (2) the hydrodynamic initial conditions used in each centrality bin. For the integrated experimental cross sections we used $\vec{\sigma}_{\text{exp}}=\{57.6$, 19, 3.72, 13.69, 16.1, 6.8, 3.27, 12.0, $14.15\}$ nb, which were obtained from the measurements of refs.~\cite{Sirunyan:2018nsz,Aaij:2014caa}. For details concerning the procedure used to obtain these cross sections, we refer the reader to Sec.~6.4 of ref.~\cite{Brambilla:2020qwo}.
The direct production cross sections appearing in Eq.~\eqref{eq:feeddown} were obtained using  $\vec{\sigma}_{\text{direct}} = F^{-1} \vec{\sigma}_{\text{exp}}$. 
 
To obtain $v_2$ in each centrality class, we computed $\langle\!\langle \cos(2(\phi-\Psi_2) \rangle\!\rangle_{i,c,p_T}$, where the average is over all bottomonium states of type $i$ produced in the corresponding centrality and transverse momentum bins and $\Psi_2$ is second-order event plane angle determined by final-state charged hadrons.  Note that $\Psi_2$ changes from event to event depending on the initial condition and fluctuations in this variable were accounted for in our computation of $v_2$.

%%%%%%%%%%%%%%%%%%%%%%%%%%%%%%%%%%%%%%%%%%%%%%%%%%%%%%%%%%%%%%
\begin{figure*}[t]
	\centering
	\includegraphics[width=0.45\linewidth]{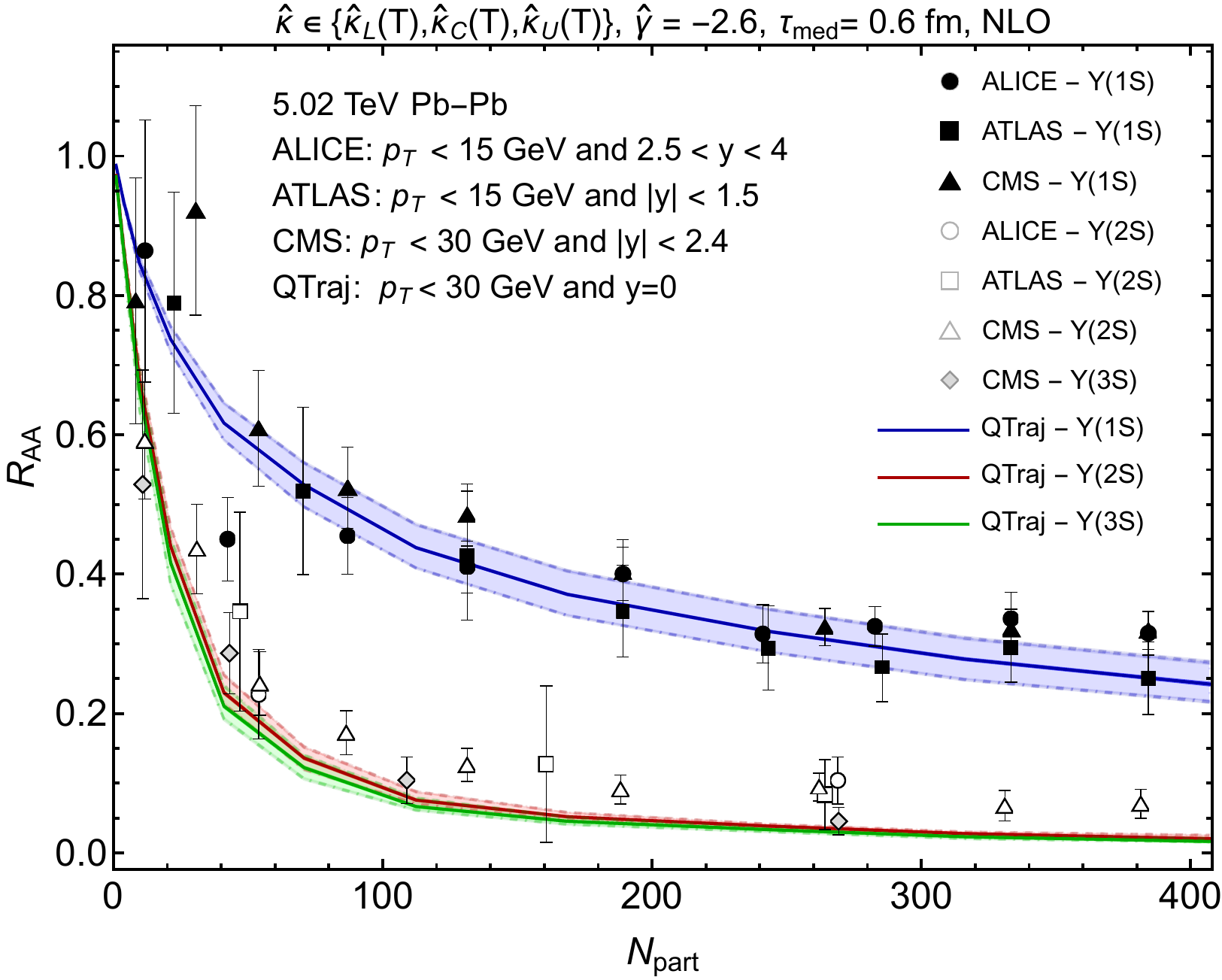} \hspace{3mm}
	\includegraphics[width=0.45\linewidth]{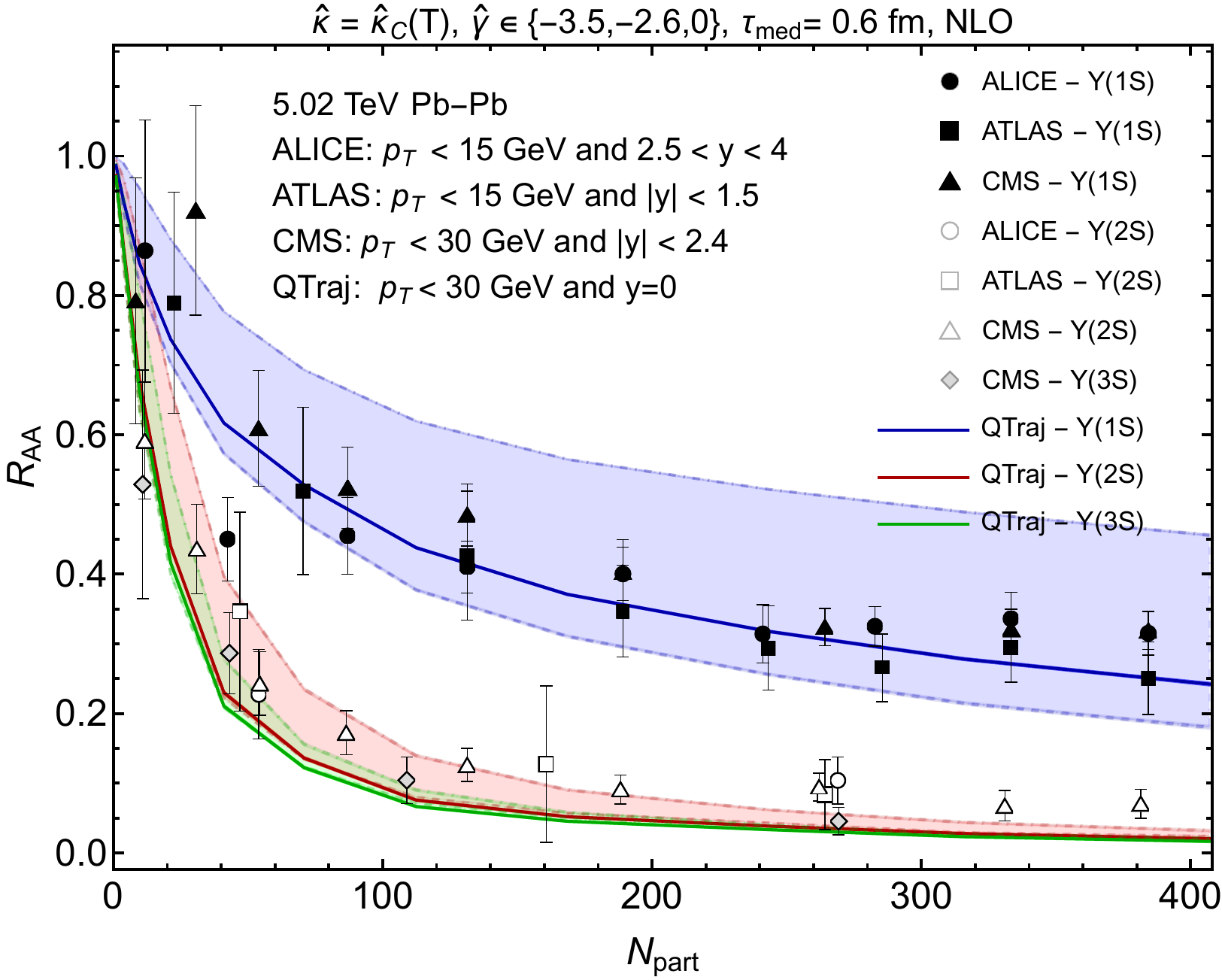}
	\vspace{-3mm}
	\caption{The nuclear suppression factor $R_{AA}$ for the $\Upsilon(1S)$, $\Upsilon(2S)$, and $\Upsilon(3S)$ states as a function of $N_{\rm part}$ obtained with IP-Glasma initial conditions.  The left panel shows variation of $\hat\kappa$ and the right panel shows variation of $\hat\gamma$.  The experimental measurements shown are from the ALICE~\cite{Acharya:2020kls}, ATLAS~\cite{ATLAS5TeV}, and CMS~\cite{Sirunyan:2018nsz,CMSupsilonQM2022} collaborations.}
	\label{fig:raavsnpart}
\end{figure*}
%%%%%%%%%%%%%%%%%%%%%%%%%%%%%%%%%%%%%%%%%%%%%%%%%%%%%%%%%%%%%%

%%%%%%%%%%%%%%%%%%%%%%%%%%%%%%%%%%%%%%%%%%%%%%%%%%%%%%%%%%%%%%
\begin{figure*}[t]
	\centering
	\includegraphics[width=0.45\linewidth]{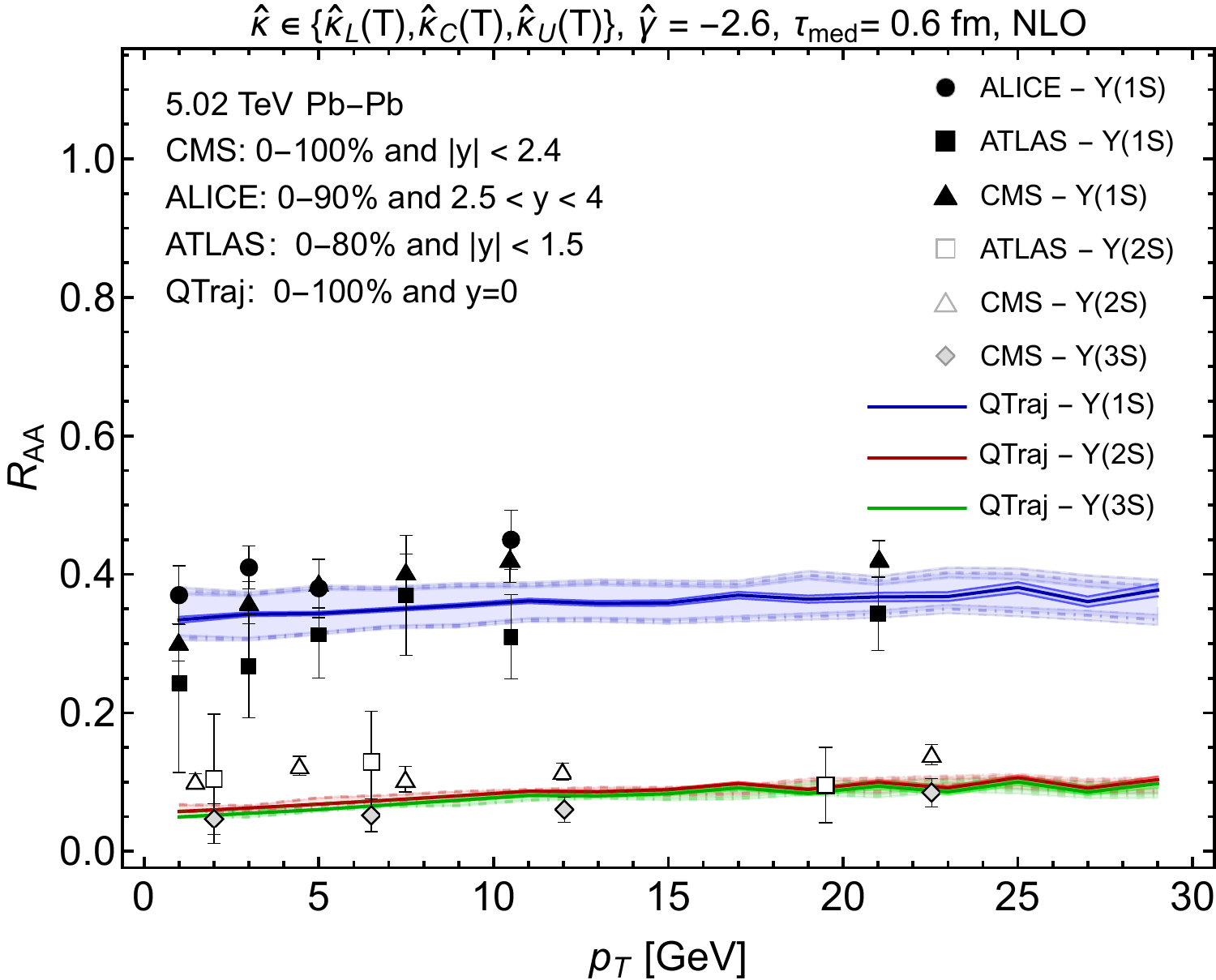} \hspace{3mm}
	\includegraphics[width=0.45\linewidth]{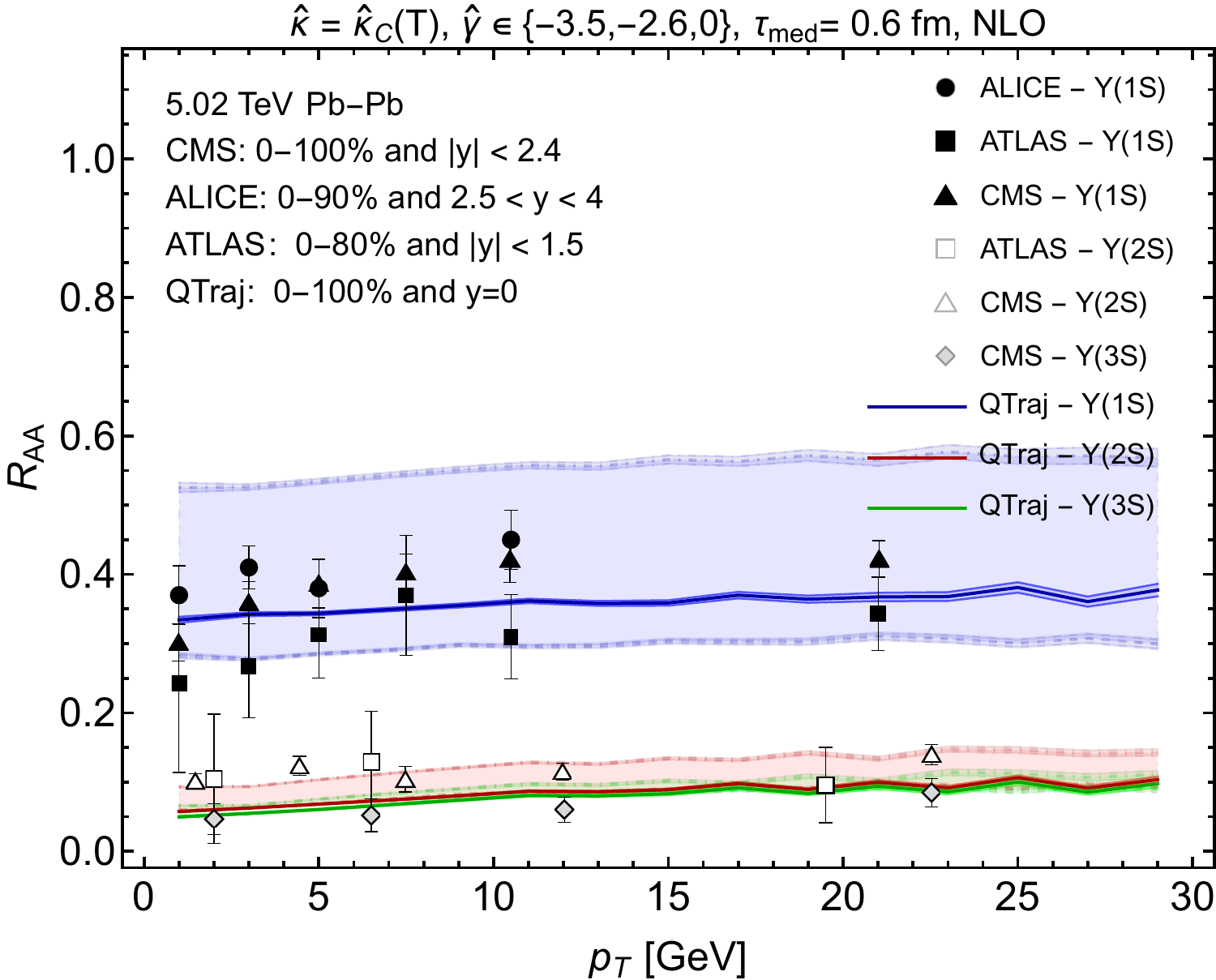}
	\vspace{-3mm}
	\caption{The nuclear suppression factor $R_{AA}$ for the $\Upsilon(1S)$, $\Upsilon(2S)$, and $\Upsilon(3S)$ states as a function of $p_T$ obtained with IP-Glasma initial conditions.  The bands and experimental data sources are the same as Fig.~\ref{fig:raavsnpart}.}
	\label{fig:raavspt}
\end{figure*}
%%%%%%%%%%%%%%%%%%%%%%%%%%%%%%%%%%%%%%%%%%%%%%%%%%%%%%%%%%%%%%

%%%%%%%%%%%%%%%%%%%%%%%%%%%%%%%%%%%%%%%%%%%%%%%%%%%%%%%%%%%%%%
\begin{figure*}[ht]
	\centering
	\includegraphics[width=0.5\linewidth]{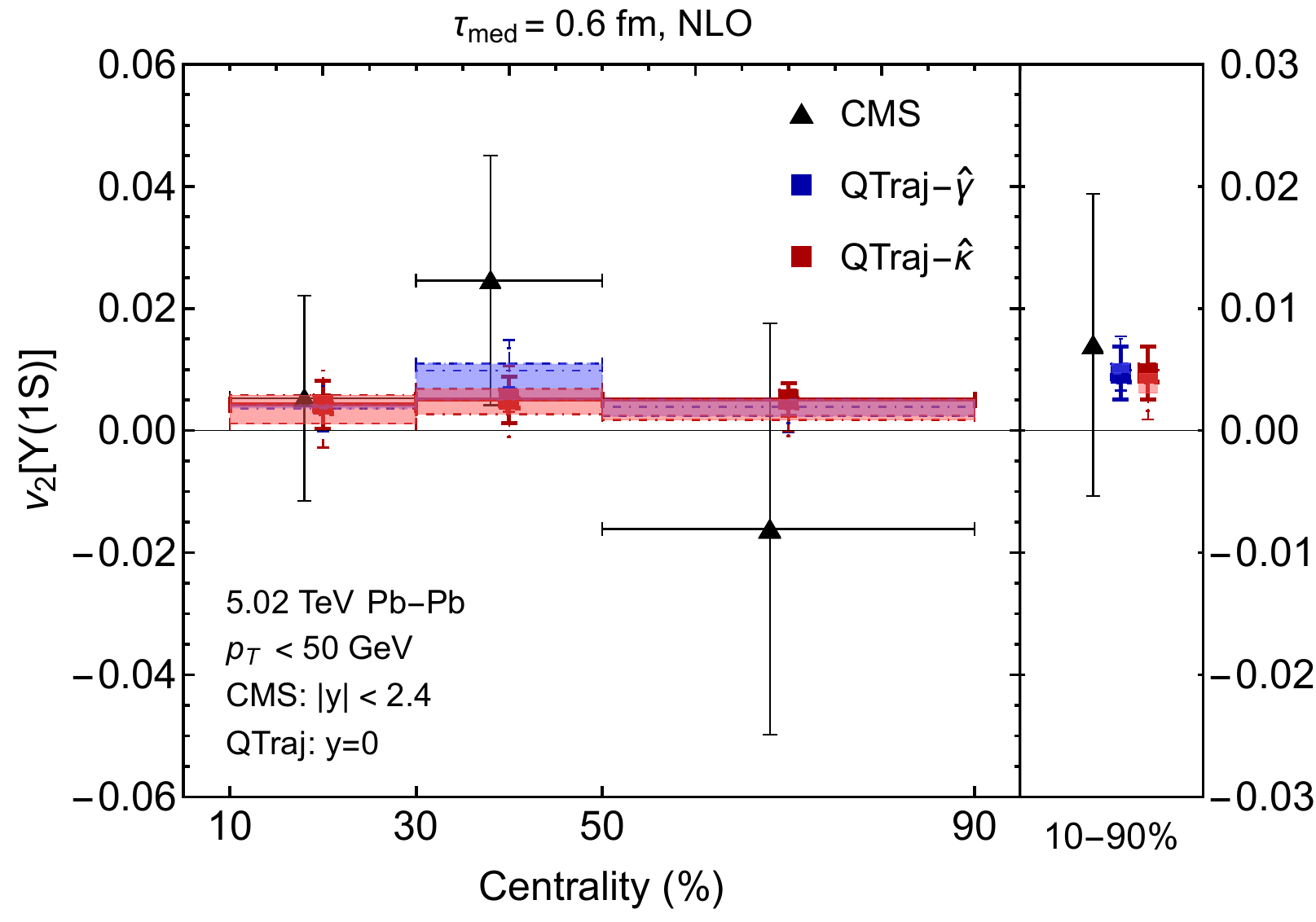}
	\hspace{5mm}
	\includegraphics[width=0.44\linewidth]{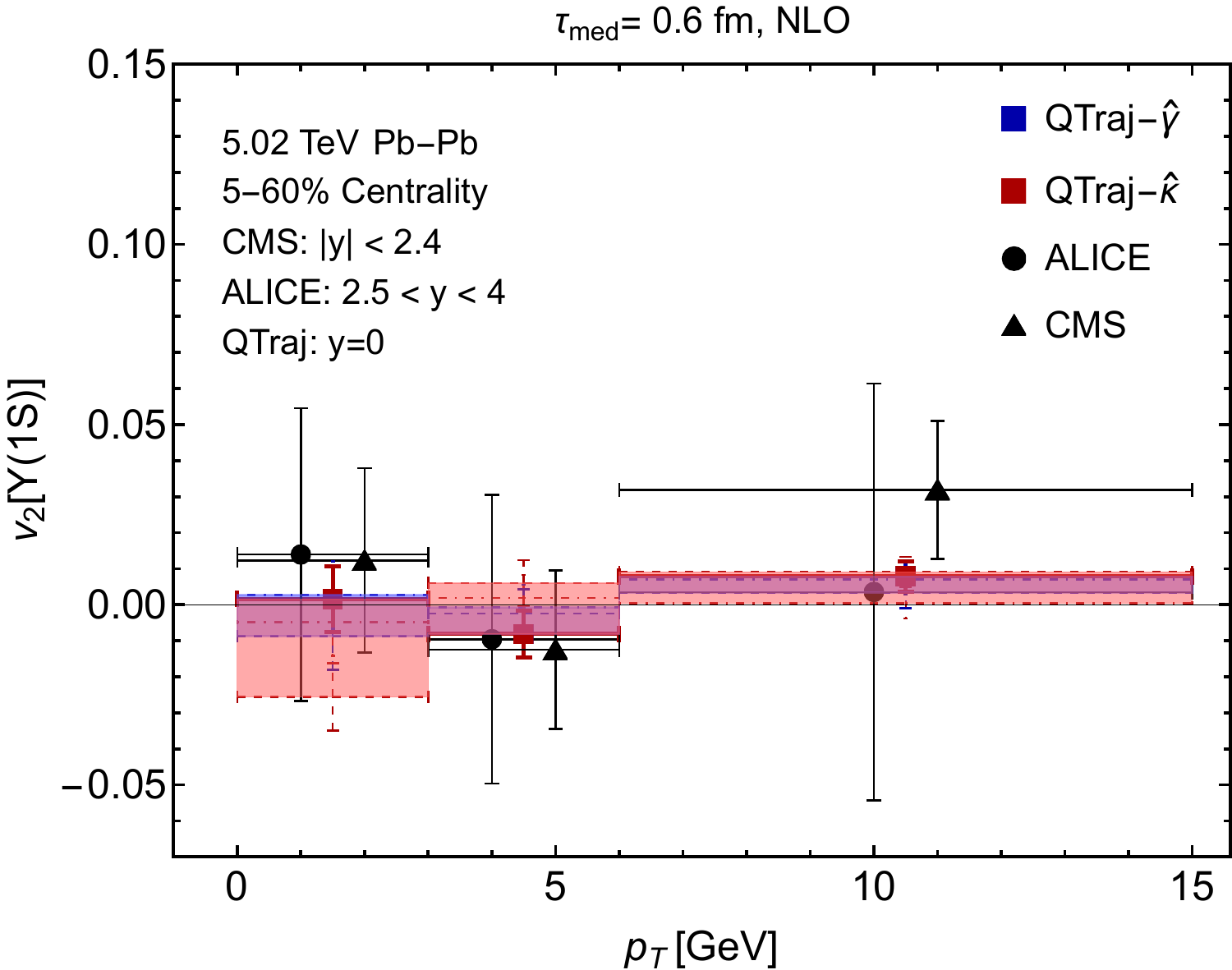}
	\vspace{-3mm}
	\caption{The anisotropic flow coefficient $v_{2}$ as a function of centrality (left) and transverse momentum (right) obtained with IP-Glasma initial conditions.  We show the $\hat\gamma$ variation in blue and the $\hat\kappa$ variation in red and compare to experimental data from the ALICE and CMS collaborations \cite{ALICE:2019pox,CMS:2020efs}.}
	\label{fig:v2}
\end{figure*}
%%%%%%%%%%%%%%%%%%%%%%%%%%%%%%%%%%%%%%%%%%%%%%%%%%%%%%%%%%%%%%

\vspace{1mm}

{\em Results} --- In Fig.~\ref{fig:raavsnpart} we present our results for $R_{AA}$ as a function of the number of participants, $N_{\rm part}$, compared to experimental data.  In both panels our statistical errors are on the order of the line width.  In the right panel, the line in the center of the bands has $\hat{\gamma} = -2.6$ which is the value that provided the best agreement with the data in Ref.~\cite{Brambilla:2022ynh}. We find, similar to Refs.~\cite{Brambilla:2020qwo,Brambilla:2021wkt,Brambilla:2022ynh}, that the variation of $R_{AA}$ with $\hat\kappa$ is much smaller than the variation with $\hat\gamma$.  This provides motivation for more constraining extractions of $\hat\gamma$ from lattice QCD studies.

As a point of reference, in the supplemental material associated with this paper, we present the same figures, but instead obtained using smooth optical Glauber initial conditions with all other parameters, etc. held fixed.  Similar results can be found in Refs.~\cite{Brambilla:2020qwo, Brambilla:2021wkt} with lower statistics.  A comparison of the results shown in Fig.~\ref{fig:raavsnpart} with those results demonstrates that the inclusion of fluctuating initial conditions results in quite small changes in the predicted $R_{AA}$ versus $N_{\rm part}$, with the differences being smaller than the systematic theoretical uncertainties associated with the variations in both $\hat\kappa$ and $\hat\gamma$.  From this figure, we see that the $\Upsilon(1S)$ $R_{AA}$ is well reproduced, however, the amount of suppression for the $\Upsilon(2S)$ is over predicted for $N_{\rm part} \gtrsim 80$.  This could be due to the fact that the pNRQCD approach used works best for the ground state which has a smaller size than the excited states.  It could also be due to the fact that in this work we did not include the effect of off-diagonal quantum jumps in the dynamical evolution, which matter more for the excited states than the ground state~\cite{Brambilla:2020qwo}.

In Fig.~\ref{fig:raavspt} we present our results for $R_{AA}$ as a function of the transverse momentum, $p_T$.  As in Fig.~\ref{fig:raavsnpart}, we see that the variation with $\hat\gamma$ is much larger than with $\hat\kappa$.  Compared to the results obtained with optical Glauber initial conditions (see the supplemental material and the figures in Ref.~\cite{Brambilla:2022ynh}), once again we find very little difference between smooth and fluctuating initial conditions.  For both smooth and fluctuating initial conditions we find that the suppression of the ground state predicted by the NLO OQS+pNRQCD approach agrees well with experimental observations, however, there is some tension between the predictions and the observed $\Upsilon(2S)$ suppression.  Despite this, we find that the framework predicts that there is a very weak dependence on $p_T$ which is consistent with experimental observations.  This should be contrasted with the $p_T$ dependence of $J/\psi$ suppression observed at LHC energies, where the experimental data indicate a strong increase in $R_{AA}$ at low $p_T$ consistent with recombination of liberated charm/anticharm quarks with other charm/anticharm quarks in the QGP \cite{ALICE:2016flj,CMS:2017uuv}.

In Fig.~\ref{fig:v2} we present our predictions for the anisotropic flow coefficient $v_{2}$ as a function of centrality in the left panel and transverse momentum in the right panel.  We compare our predictions with experimental data from the ALICE and CMS collaborations \cite{ALICE:2019pox,CMS:2020efs}.  As the left panel of Fig.~\ref{fig:v2} demonstrates, the NLO OQS+pNRQCD framework predicts a rather flat dependence on centrality, with the maximum $v_2$ being on the order of 1\%.  In the right portion of the left panel, we present the results integrated over centrality in the 10-90\% range as two points that include the observed variations with $\hat\kappa$ and $\hat\gamma$, respectively.  Note, importantly, that the scale of the right portion of the left panel is different from the left portion of this panel.  The size of the error bars reflects the statistical uncertainty associated with the double average over initial conditions and physical trajectories and the light shaded regions correspond to the uncertainty associated with the variation of $\hat\kappa$ and $\hat\gamma$, respectively.  

Considering both variations, we find that when integrated in the 10-90\% centrality interval and \mbox{$p_T < 50$ GeV}, the $v_2$ of the $\Upsilon(1S)$ is \mbox{$v_2[1S] = 0.005 \pm 0.002 \pm 0.001$}, with the first number corresponding to the statistical uncertainty and the second the systematic uncertainty associated with the variation of both $\hat\kappa$ and $\hat\gamma$. Within statistical uncertainties, this is consistent with the results reported in Refs.~\cite{Bhaduri:2020lur,Islam:2020bnp,Brambilla:2021wkt} and also those presented in the supplementary material, where optical Glauber initial conditions were used.  Comparing with the supplementary plots, which is an apples-to-apples comparison of NLO OQS+pNRQCD with fluctuating and smooth initial conditions, there are hints of a slight decrease in the integrated $v_2[1S]$, however, the decrease is within our statistical uncertainty.  Finally, turning to the right panel of Fig.~\ref{fig:v2} we see that the dependence of $v_2[1S]$ on transverse momentum is rather flat, however, we note that at low momentum there is a stronger dependence on $\hat\kappa$, which could help to further constrain this parameter if more precise experimental data is made available.

\vspace{1mm}

{\em Conclusions} --- In this paper we presented the first results concerning the impact of fluctuating hydrodynamic initial conditions on bottomonium production within a dynamical open quantum systems approach.  The complex Hamiltonian used for the quantum evolution is accurate to next-to-leading order in the binding energy over temperature, having been recently obtained in Refs.~\cite{Brambilla:2022ynh}.  Due to the computational demand of averaging over both bottomonium trajectories and fluctuating initial conditions, herein we have ignored the effect of dynamical quantum jumps, which have been shown to be small in Refs.~\cite{Brambilla:2020qwo,Brambilla:2021wkt}. In a forthcoming longer paper, we will present predictions for the elliptic flow of 2S and 3S excited states, along with predictions for higher-order anisotropic flow coefficients such as $v_3$ and $v_4$ of all states using the methodology introduced in this paper.  

Looking to the future, it will be important to determine the effect of off-diagonal quantum jumps on both $R_{AA}$ and $v_n$.  Given sufficient computational resources, this can be accomplished using the existing quantum trajectories code.  It would also be interesting to see if full 3D fluctuating initial conditions have any impact on the rapidity dependence of these observables.  Finally, one outstanding theoretical uncertainty of our work is the effect of the center of mass velocity of the quarkonium state being different than the local flow velocity of the QGP.  This effect should be more pronounced when including fluctuating initial conditions, since the flow velocity is more non-uniform, however, it has not yet been included in phenomenological models, even in the case of smooth initial conditions.

\vspace{1mm}

{\em Acknowledgements} ---
H.A. was supported by the Deanship of Scientific Research at Umm Al-Qura University under Grant Code 22UQU4331035DSR02.
J.B. and M.S. were supported by U.S. DOE Award No.~DE-SC0013470 and C.S. by U.S. DOE Award No. DE-SC0021969.

%%%%%%%%%%%%%%%%%%%%%%%%%%%%%%%%%%%%%%%%%%%%%%%%%%%%%%%%%%%%%%
\bibliography{qtraj-fluc}
%%%%%%%%%%%%%%%%%%%%%%%%%%%%%%%%%%%%%%%%%%%%%%%%%%%%%%%%%%%%%%

\clearpage

\begin{widetext}

\section*{Supplemental material}

In this supplement, we provide the results of runs using non-fluctuating optical Glauber initial conditions.  The hydrodynamical runs used here are precisely the same as those used in Ref.~\cite{Brambilla:2022ynh}, which made use of the anisotropic hydrodynamics formalism~\cite{Martinez:2010sc,Alqahtani:2017mhy,Alqahtani:2020paa}.  The simulation parameters for the evolution of the bottomonium wave function and the number of trajectories sampled were the same as were used for the IP-Glasma fluctuating initial condition runs presented in the main body.  Note that, compared to Ref.~\cite{Brambilla:2022ynh}, we considered 200,000 trajectories instead of 20,000 trajectories. 

In Fig.~\ref{fig:raavsnpart-smooth}, we present the smooth hydrodynamical initial condition results for $R_{AA}$ as a function of $N_{\rm part}$.  We note that the central value and bands found with 200,000 trajectories are slightly lower than those obtained with 20,000 trajectories in Ref.~\cite{Brambilla:2022ynh}.  We have confirmed that this change is consistent with the statistical uncertainties associated with the lower number of trajectories used in Ref.~\cite{Brambilla:2022ynh}.  In Fig.~\ref{fig:raavspt-smooth}, we present our results obtained for $R_{AA}$ as a function of $p_T$ using optical Glauber initial conditions.  Finally, in Fig.~\ref{fig:v2-smooth}, we present our results obtained for $v_2$ as a function of centrality and $p_T$ using optical Glauber initial conditions.

%%%%%%%%%%%%%%%%%%%%%%%%%%%%%%%%%%%%%%%%%%%%%%%%%%%%%%%%%%%%%%
\begin{figure}[ht]
	\centering
	\includegraphics[width=0.45\linewidth]{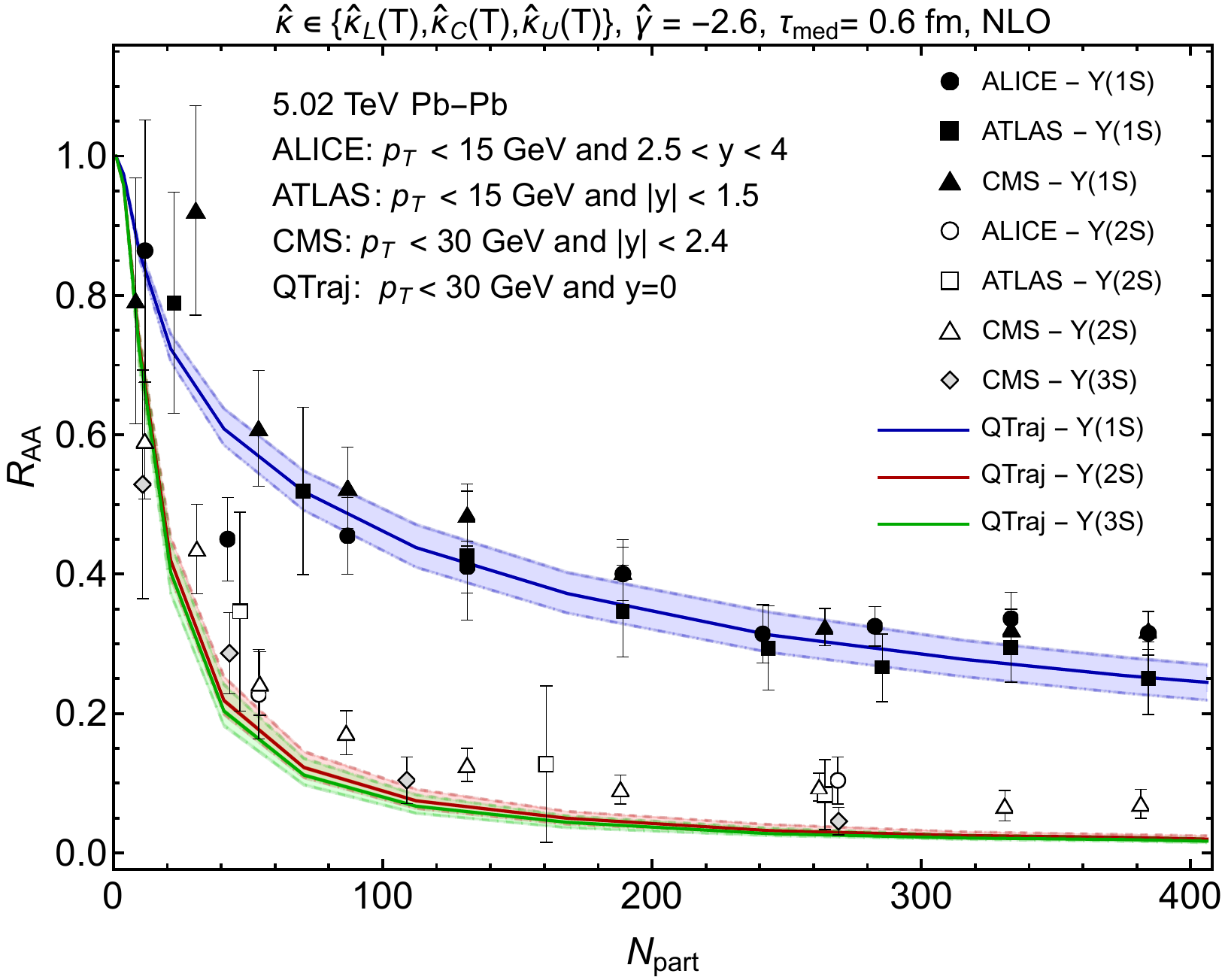} \hspace{5mm}
	\includegraphics[width=0.45\linewidth]{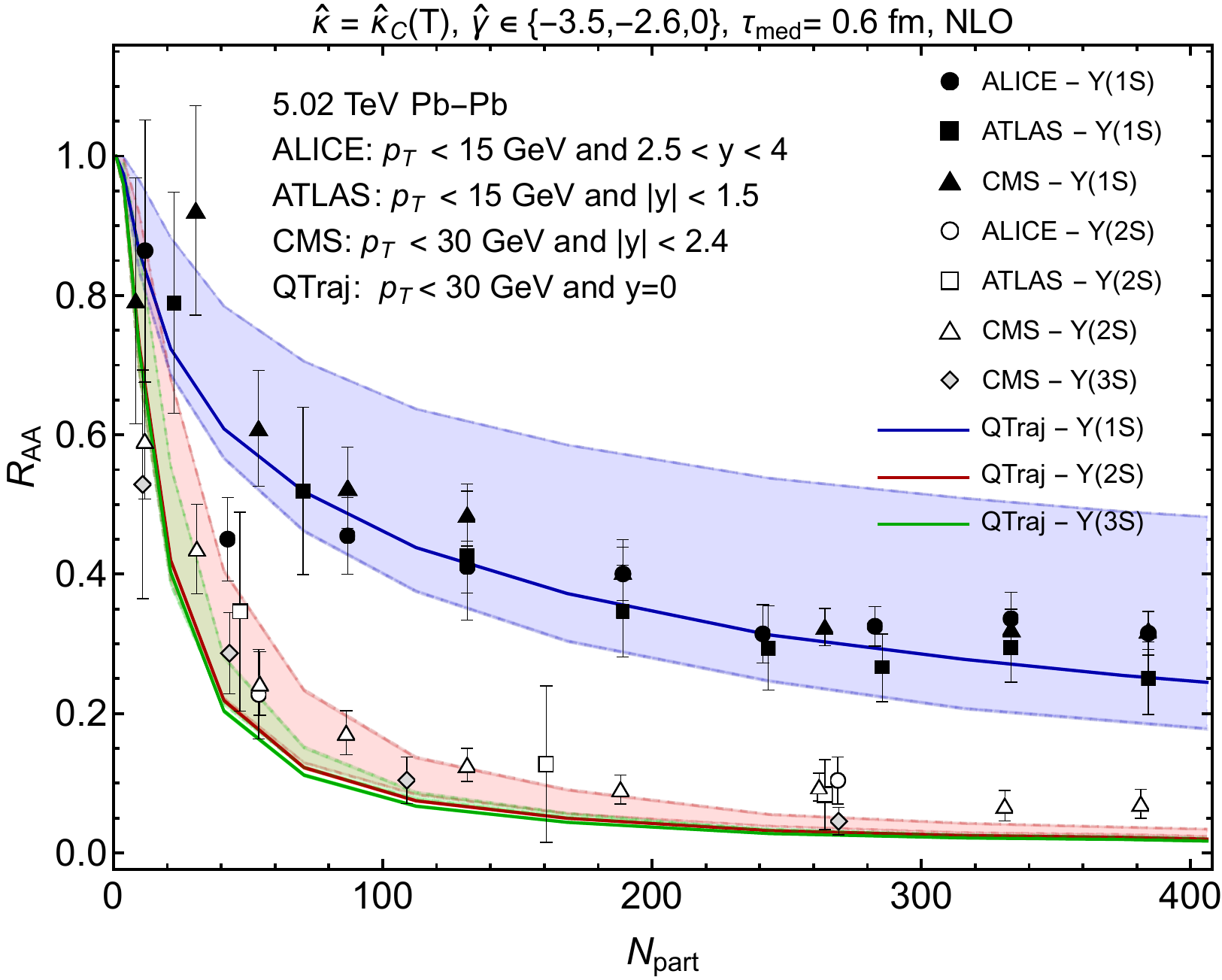}
	\vspace{-3mm}
		\caption{The nuclear suppression factor $R_{AA}$ for the $\Upsilon(1S)$, $\Upsilon(2S)$, and $\Upsilon(3S)$ states as a function of $N_{\rm part}$ obtained with optical Glauber initial conditions.  The left panel shows variation of $\hat\kappa$ and the right panel shows variation of $\hat\gamma$.  The experimental measurements shown are from the ALICE~\cite{Acharya:2020kls}, ATLAS~\cite{ATLAS5TeV}, and CMS~\cite{Sirunyan:2018nsz,CMSupsilonQM2022} collaborations.}
	\label{fig:raavsnpart-smooth}
\end{figure}
%%%%%%%%%%%%%%%%%%%%%%%%%%%%%%%%%%%%%%%%%%%%%%%%%%%%%%%%%%%%%%

%%%%%%%%%%%%%%%%%%%%%%%%%%%%%%%%%%%%%%%%%%%%%%%%%%%%%%%%%%%%%%
\begin{figure*}[ht]
	\centering
	\includegraphics[width=0.45\linewidth]{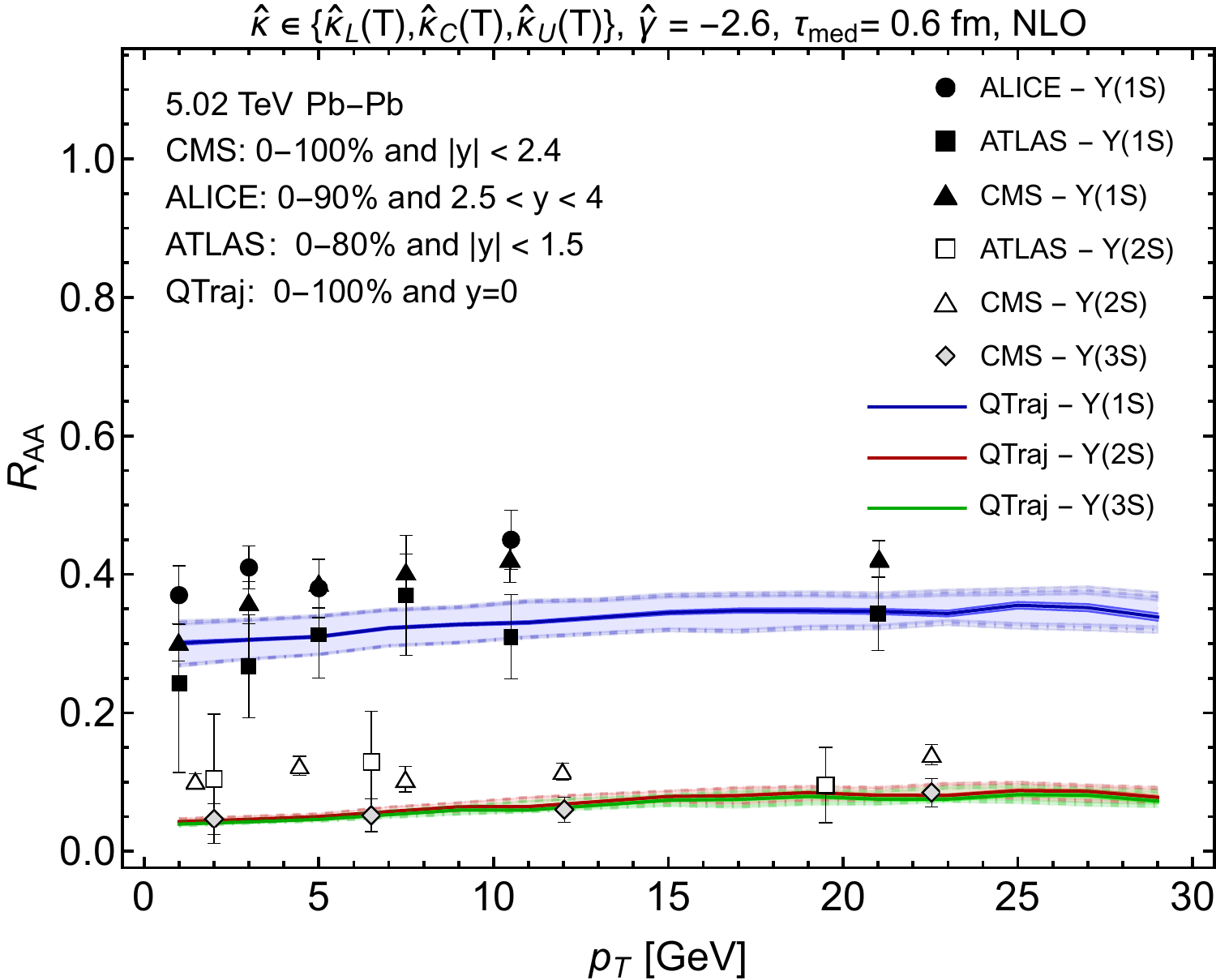} \hspace{3mm}
	\includegraphics[width=0.45\linewidth]{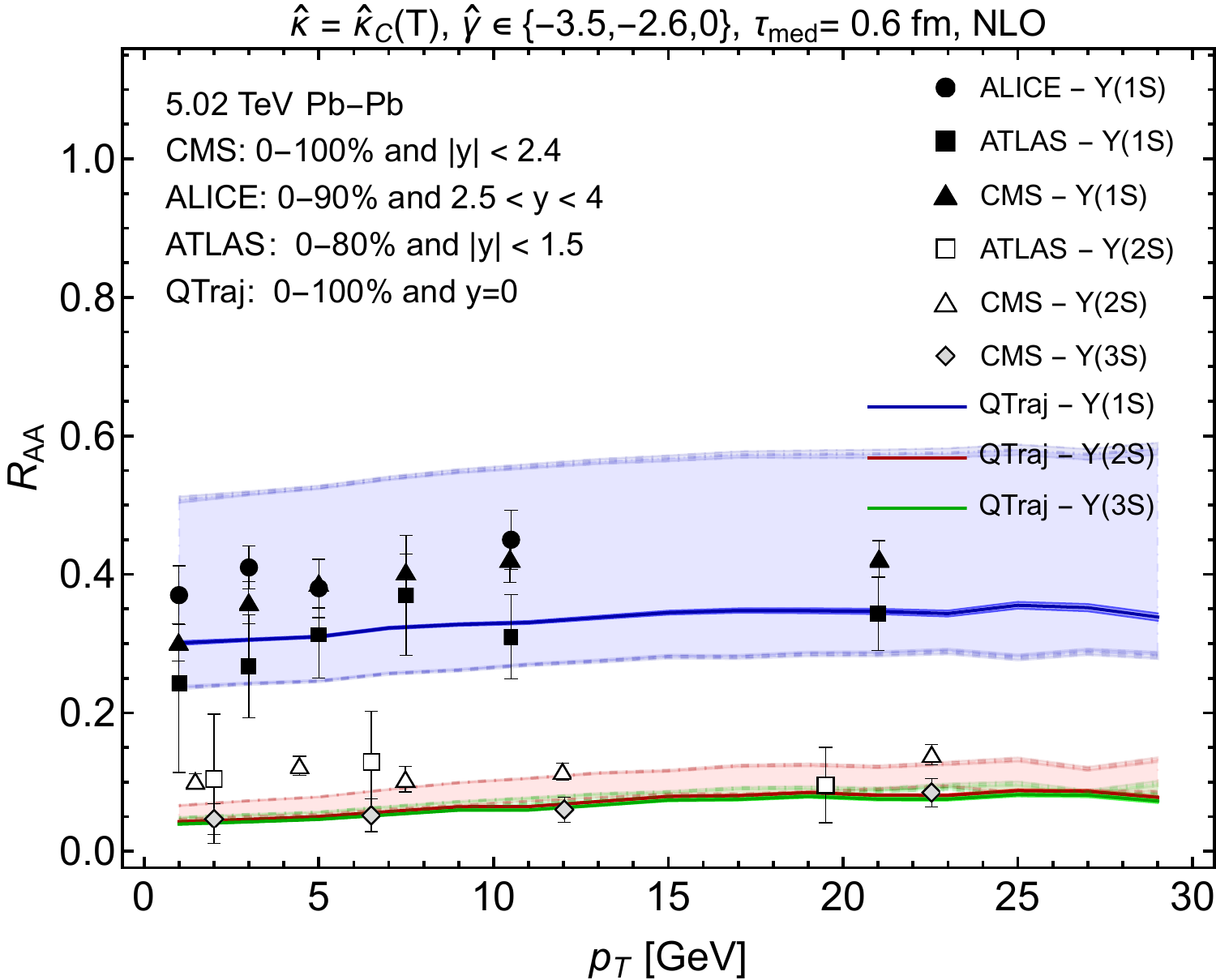}
	\vspace{-3mm}
	\caption{The nuclear suppression factor $R_{AA}$ for the $\Upsilon(1S)$, $\Upsilon(2S)$, and $\Upsilon(3S)$ states as a function of $p_T$ obtained with optical Glauber initial conditions.  The bands and experimental data sources are the same as Fig.~\ref{fig:raavsnpart-smooth}.}
	\label{fig:raavspt-smooth}
\end{figure*}
%%%%%%%%%%%%%%%%%%%%%%%%%%%%%%%%%%%%%%%%%%%%%%%%%%%%%%%%%%%%%%

%%%%%%%%%%%%%%%%%%%%%%%%%%%%%%%%%%%%%%%%%%%%%%%%%%%%%%%%%%%%%%
\begin{figure}[ht]
	\centering
	\includegraphics[width=0.475\linewidth]{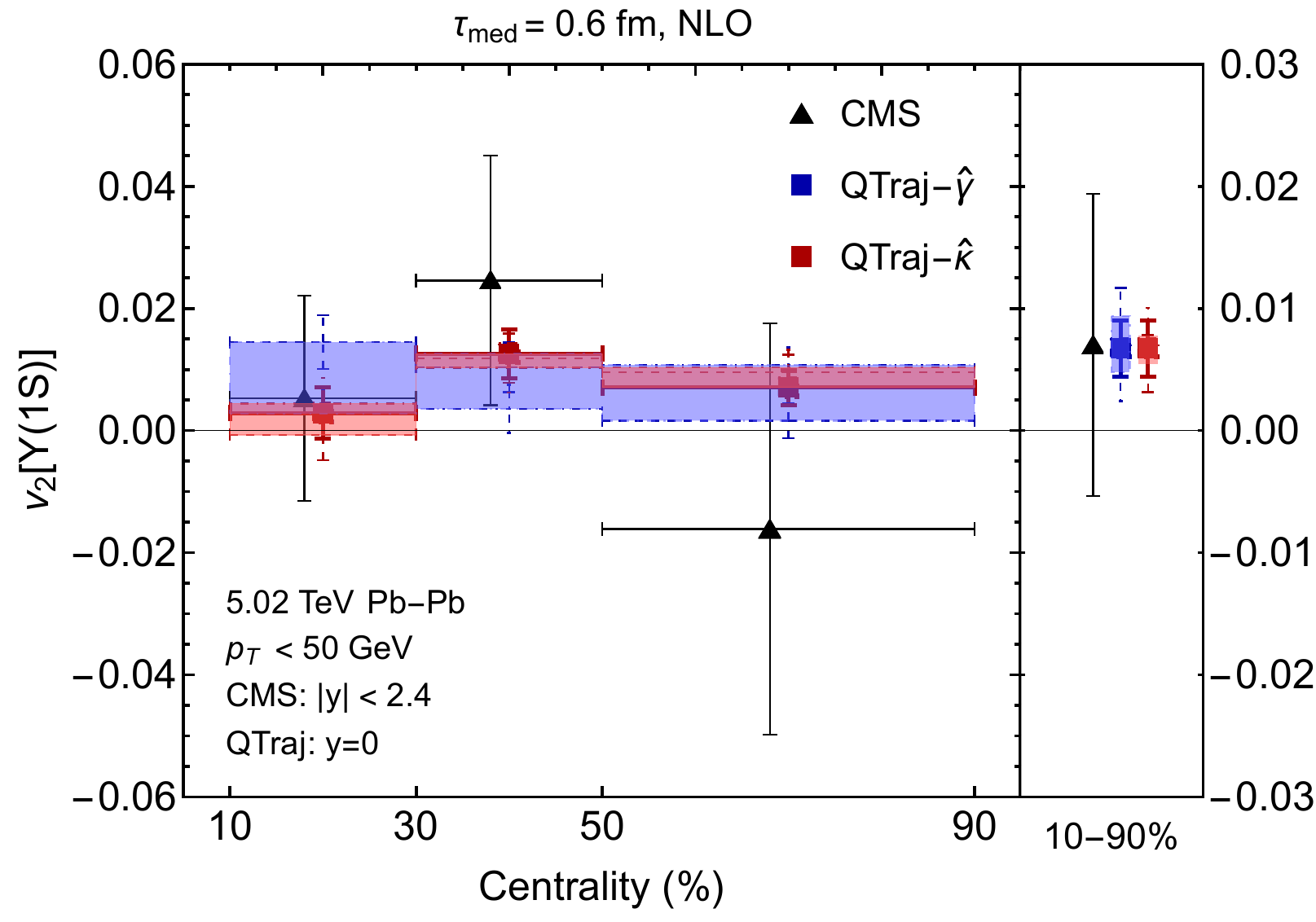} \hspace{3mm}
	\includegraphics[width=0.425\linewidth]{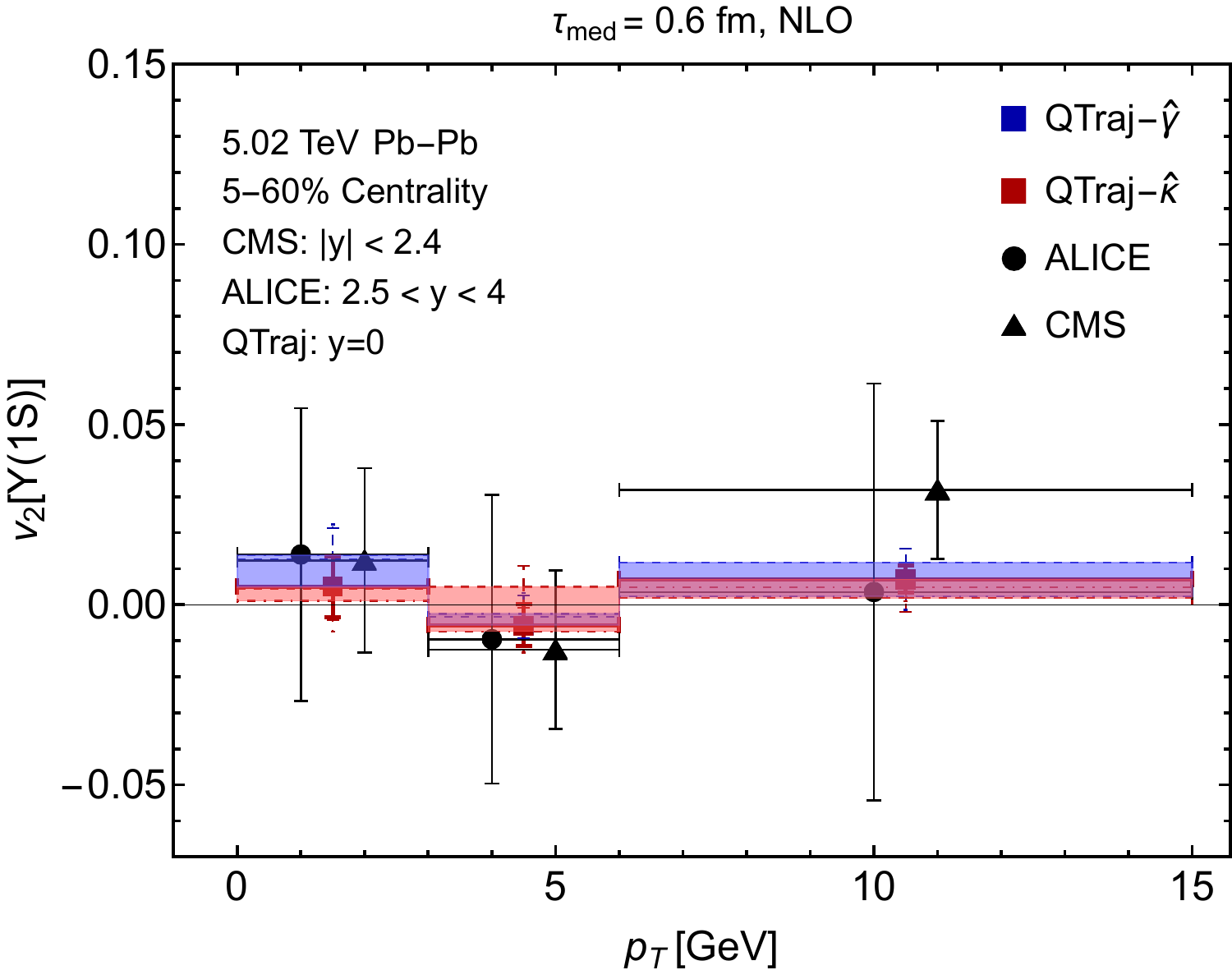}
	\vspace{-2mm}
	\caption{The anisotropic flow coefficient $v_{2}$ as a function of centrality (left) and transverse momentum (right) obtained with optical Glauber initial conditions.  We show the $\hat\gamma$ variation in blue and the $\hat\kappa$ variation in red and compare to experimental data from the ALICE and CMS collaborations \cite{ALICE:2019pox,CMS:2020efs}.}
	\label{fig:v2-smooth}
\end{figure}
%%%%%%%%%%%%%%%%%%%%%%%%%%%%%%%%%%%%%%%%%%%%%%%%%%%%%%%%%%%%%%

\end{widetext}

\end{document}